# Regenerated Cellulose Fiber Solar Cell

Michael Ebner[1], Robert Schennach[1], Huei-Ting Chien[1], Claudia Mayrhofer[2], Armin Zankel[3], Bettina Friedel[1]*

[1] *Institute of Solid State Physics, Graz University of Technology, Petersgasse 16, 8010 Graz, Austria*

[2] *Graz Centre for Electron Microscopy, Steyrergasse 17, 8010 Graz, Austria*

[3] *Institute for Electron Microscopy and Nanoanalysis, Graz University of Technology, Steyrergasse 17, 8010 Graz, Austria*

* Corresponding Author email: bfriedel@tugraz.at

Abstract

Wearable electronics and smart textiles are growing fields in the cause to integrate modern communication and computing tools into clothing instead of carrying around smart phones and tablets. Naturally, this also requires power sources to be integrated in textiles. In this paper, a proof-of-concept is presented in form of a photovoltaic cell based on commercially available viscose fiber. This has been realized using a silver nanowire network around the viscose fiber to establish electrical contact and a photoactive coating using the standard workhorse among organic thin film solar cells, a blend of poly(3-hexylthiophene) (P3HT) and phenyl-$C_{61}$-butyric acid methyl ester (PCBM). Structure and performance of single fiber devices demonstrate their feasibility and functionality. The applied materials and methods are compatible to solution processing therewith qualifying for potential roll-to-roll large-scale production.

*Regenerated Cellulose Fibers, Organic Photovoltaic Cells, Flexible Electronics, Wearable Electronics, Conducting Cellulose Fibers*

# Introduction

Nowadays, people increasingly rely on portable electronics from mobile phones to tablets and a whole range of other gadgets. Manufacturers are now trying to integrate the technology directly into clothing. This concept is known as smart textiles (Service 2003; Weng et al. 2016). However, all these modern helpers require electrical energy to work. At present, rechargeable batteries are the main solution to provide the power needed. But these have two main disadvantages limiting their practicability for wearable electronics. First, their storage capacity is limited and second, their weight increases with capacity. One possible solution to this problem is energy conversion. Therein an entire field of research is dedicated to the transformation of different human daily life energy sources (e.g. light, body heat, movement) into usable electrical power (Weng et al. 2016).

Light, as an abundant source of energy, motivated various reported concepts for photovoltaic devices that are compatible with textiles (Weng et al. 2016). In one approach, a textile-based solar cell has been realized by stacking textile electrodes of woven metal, carbon and dye-decorated titania wires, finalized by adding an electrolyte to make a dye-sensitized solar cell (e.g. (Pan et al. 2014)). However, the practical handling of the electrolyte in the final fabric is problematic. In another report, a standard inverted organic photovoltaic device structure was deposited onto a gold-decorated textile electrode and stitched onto a fabric, to be connected by conducting fibers to potential appliances (Lee et al. 2014). There, the durability of a stitched-on device during wear is the limiting factor. There have been also attempts to fabricate photovoltaic fibers to enable direct integration of the solar cell into the fabric. For this latter approach versatile concepts have been presented. For example (Qiu et al. 2014) reported a perovskite solar cell assembled onto a flexible stainless steel wire. Even a dye-sensitized solar cell was transferred into a fiber design by (Chen et al. 2012) by using intertwined strands of carbon nanotube (CNT) fiber with and without titania/dye coating. An organic solar cell fiber using the common conjugated polymer P3HT and fullerene derivative PCBM was presented by (Zhang et al. 2014) built on a titanium wire electrode, stranded with a CNT fiber counter electrode. One thing these approaches have in common is the metal or carbon based conductive fiber as center electrode. Another approach was chosen by (Friedel et al. 2009) who built an organic-inorganic hybrid solar cell based on differently doped single-crystalline silicon carbide (SiC) fibers, which are optically transparent and depending on the dopant concentration conductor or semiconductor. As the latter, SiC was used as the acceptor in combination with the conjugated polymer donor P3HT deposited on the fiber.

The main drawback of most of the mentioned fiber-based photovoltaic device designs is the limited mechanical durability of the metallic, spun-CNT, ceramic or ceramic-decorated core-fibers to bending and stretching, compared to common textile fibers. Therefore, this paper presents a photovoltaic fiber based on a regenerated cellulose (viscose) fiber, which is readily found in common fabrics. This viscose fiber is decorated with a network silver nanowires (AgNW) to form the conductive electrode, followed by a coating of organic semiconductors for the photoactive layer, finalized by a one-sided metal counter electrode. In this paper, we focused on materials and methods enabling scalability e.g. adaptable to role-to-role (R2R) processing, as outlined in (Bedeloglua et al. 2009).

In the following, the architecture, production and electronic properties of these viscose fiber-based solar cells will be presented and discussed.

## Experimental

Materials

Hollow viscose fibers with a linear mass density of 2.1 dtex (equivalent to a diameter of approx. 20 µm) and a length of about 40 mm were used in this study and provided by Kelheim Fibres GmbH (Kelheim, Germany). Silver nanowire (AgNW) suspension was supplied by Sigma-Aldrich (0.5 wt% in isopropanol) and the dispersed AgNWs had a nominal average diameter of 115 nm and lengths between 20 and 50 µm. Further, anhydrous chlorobenzene as solvent for the organic semiconductors was purchased from Sigma-Aldrich. The PEDOT:PSS (Clevios™ PVP CH 8000), was obtained from Heraeus-Clevios. Regioregular P3HT (EE-4002, RR 93 %, $M_W$ 70,000 gmol$^{-1}$) and PCBM (99%) were bought from Rieke Metals Inc. (US) and Ossila Ltd. (UK), respectively. All substances mentioned were used without further purification.

Preparation of photovoltaic fibers

For better handling under laboratory conditions, four viscose fibers at a time were fixed equidistantly with their two ends to two separate glass plates, as shown in Figure 1a, allowing access to the fiber without influence of a foreign contact surface. Nail polish was used as adhesive between viscose fiber and glass, which differently to other regular glues does not tend to creep and spread along cellulosic fibers, forming an unwanted coating (Fischer et al. 2014).

The following manufacturing procedure is based on (Kopeinik et al. 2015). To form a conductive surface for an electrode, the silver nanowires were deposited on the viscose fibers via dip-coating (SDI company, Nano-DIP ND-0407) at withdrawal speed between 11 and 66 µm s$^{-1}$ from isopropanol suspension and allowed to dry for 1 hour at room temperature. The viscose-AgNW fibers were treated with oxygen plasma (Diener Electronics, FEMTO Low Pressure Plasma System) at 100 W for 10 min and 80 cm$^3$/min O$_2$ flow, to facilitate wetting for follow-on coatings. For deposition of the hole-conduction interlayer, PEDOT:PSS suspension was sonicated for 30 minutes and 100 µl were drop-casted onto the freshly plasma-treated fiber, followed by drying in flowing argon at elevated temperature (200°C, 30 min).

The active layer was applied in argon atmosphere by repeated drop-casting from hot (70°C) solution of 1:2 P3HT:PCBM in chlorobenzene (17 mg/mL + 34 mg/mL). The layer was applied in three steps of 70 µl, 100 µl and 100 µl, with 15 minutes drying at 70° C after each step and final drying at room temperature for at least 1 hour.

Finally, the outer cathode of aluminum (~ 100 nm) was deposited via thermal evaporation on one side of the fiber.

Characterization

Visual investigation of the fibers was performed with an Olympus BX 51 optical microscope with integrated digital camera. Electrical characterization of the AgNW-coated viscose fiber and photovoltaic device characterization were carried out with electrical microprobes in argon atmosphere, using a Keithley 2636A source-measure unit. For photocurrent measurements, current–voltage characteristics were acquired under white light illumination. In absence of standard AM1.5G (100 mW/cm$^2$) light source (solar simulator) access in the glovebox, an LED white light source of color temperature of 7000 K (Osram LED "cool daylight", arrangement of 3 LEDs ) with a light intensity of 0.54 mW/cm$^2$, was used.

The surface of AgNW-coated viscose fibers was investigated by scanning electron microscopy (SEM) imaging with the Everhart–Thornley detector (ETD) of the high resolution SEM Zeiss Ultra 55. For transmission electron microscopic (TEM) evaluation of cross-sections of completed photovoltaic fibers, the hollow viscose fiber with their coatings of AgNWs, PEDOT:PSS and photoactive composite were sputter-coated with a thin gold layer and then embedded in cyanoacrylate between acrylic glass platelets and sliced using an ultra-microtome (UC6, Leica Microsystems) equipped with an 35° diamond knife (Diatome AG, Switzerland). The clearance angle was 4°, the nominal feed was generally set at 70 nm and the cutting speed was chosen at 0.6 mm/s. Theses sections were then transferred (Perfect Loop, Diatome) to a 75 mesh grid coated with formvar. The sliced cross sections of the photovoltaic fiber were examined using a Tecnai T12 (FEI Europe) transmission electron microscope with $LaB_6$ field emission cathode, operated at 120 kV in bright-field mode, with a 7.5 mrad objective aperture.

## Results and Discussion

The principal design of the suggested photovoltaic viscose fibers is as follows (see Figure 1b): First, the viscose fiber is coated with a continuous but flexible conductive material to create the flexible inner electrode.

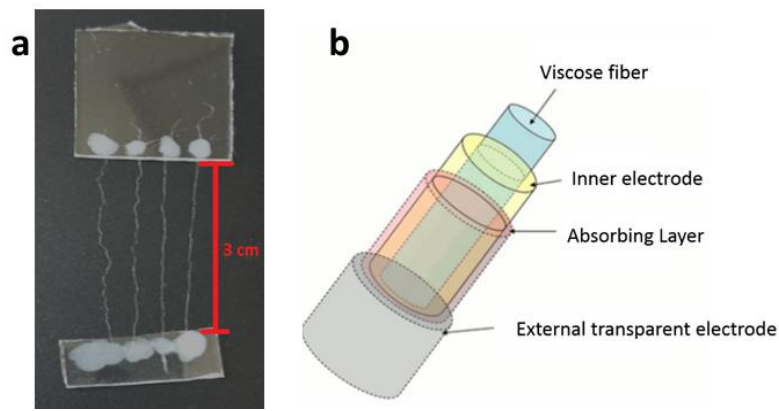

Figure 1: (a) Photograph of preparation arrangement of four viscose fibers fixed to glass slides. (b) Sketch of the principal lineup of the photovoltaic fiber based on a viscose fiber.

A follow-on hole-conductor coating will act as a buffer to prevent short circuits, lead to selectivity of the electrode and diminish charge extraction barriers. This is followed by photoactive (absorbing) layer consisting of solution processable organic semiconductors. The final layer is the external transparent electrode. Suitable solution-processable conductive materials for the outer electrode are sol-gel derived transparent conducting oxides (TCO), as found in literature (Granqvist 2007; Hilgendorff et al. 1998; Kisailus et al. 2006; Morgenstern et al. 2011) or graphene (Wei et al. 2010; Zhu et al. 2010).

In the present proof-of-concept, the devices have the following configuration: Viscose/AgNW/PEDOT:PSS/P3HT:PCBM/Al. The choice of AgNWs for the inner electrode has multiple reasons: First, their convenient solution-processibility; second, the good adhesion and network-building properties on cellulose surfaces, as has been shown earlier by (Kopeinik 2015); and third the immense benefit of the resulting large metallic surface area, which is considerably higher than if merely a smooth continuous coating of silver was applied, generating a larger contact area for charge-transfer from the semiconductor, thus enabling a larger current-output per area (of the fabric). PEDOT:PSS as hole-conductor and photoactive blends of P3HT as donor with PCBM as acceptor are standard materials used in conventional organic thin film solar cells(Dang, Hirsch, Wantz 2011). For the outer electrode in our fiber devices, a one-sided evaporated aluminum cathode was chosen for simplicity reasons. This means that here a common model system from planar organic photovoltaic devices (Dang, Hirsch, Wantz 2011) is adapted for application on a 1D-device based on regenerated cellulose fiber, decorated with silver nanowires, as the electrode. For planar devices with these materials and architecture, power conversion efficiencies of up to 5 % were reported (Dang, Hirsch, Wantz 2011).

In the SEM image of the viscose fibers used in this study (Figure 2) the hollow tubular structure and corrugated surface can be clearly seen. In principle, any other type of viscose fiber with a more or less smooth outer surface could be used, too.

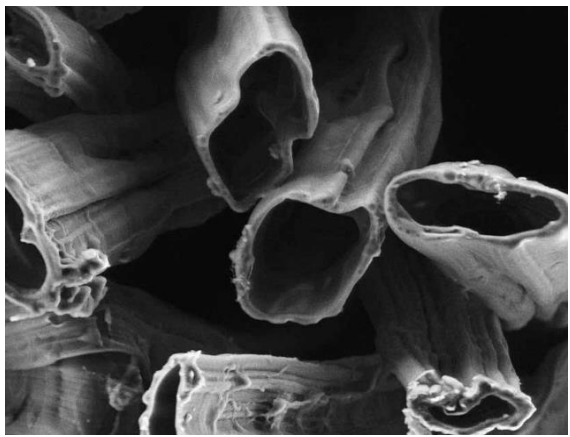

Figure 2: SEM image of the regenerated cellulose fibers used in this study. The fiber diameter is approximately 20 µm. Picture courtesy of Kelheim Fibres GmbH.

To the bare eye the viscose fibers coated with AgNWs appear white to light grey, depending on the coating conditions affecting their coverage. Under the optical microscope (Figure 3a) the transparent viscose fiber and the attached silver nanowires are already well distinguishable by the darker appearing fiber and the bright lines on the surface, reflections of the silver nanowires. However, a much more detailed picture of the adhered AgNWs on the viscose fiber surface can be seen in the SEM image in Figure 3b. Though the hollow fiber structure is obviously collapsed (due to capillary forces and vacuum), therefore appearing flat and deformed instead of tubular, the dense network of tight-fitting, closely adsorbed nanowires is clearly visible.

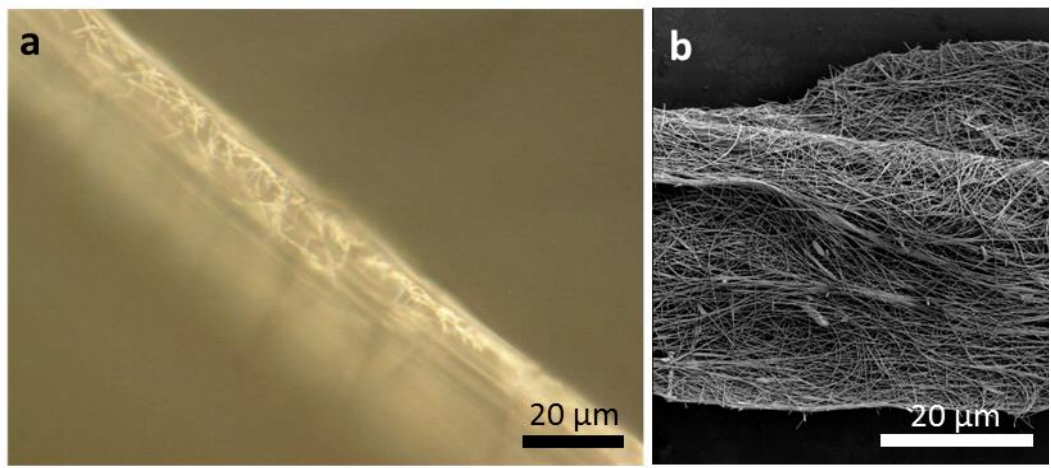

Figure 3: Cellulose fiber after dip coating in silver nanowire solution in (left) an optical microscope and (right) a SEM image.

Silver nanowires are known to enable very small bending radii, with bending angles as low as 7° to 20° observed for comparable AgNWs (Wu et al. 2006). This fact explains the observation that hardly any silver nanowires stick out at the edges of the viscose fiber, as the adhesion force to the cellulose is stronger than the elasticity of the nanowire (McDowell, Leach, Gall 2008). However, even few upstanding AgNW can easily lead considerable leakage currents or even short circuits in the device, if they cannot be sufficiently covered by the thin organic semiconductor coating. For this reason, the yield of functioning photovoltaic fibers under non-optimized laboratory conditions was at 4% (out of 80 devices). This problem could be solved by using viscose fibers that do not collapse during processing, as they would provide higher probability to achieve homogeneous semiconductor surface coatings. The conductivity of the silver nanowire decorated viscose fiber depends on its surface coverage. This amount can be tuned by the solid content of the silver nanowire suspension and the withdrawal speed in the dip-coating process. The lowest resistivity of about 30 Ω was achieved with a concentration of 0.5 wt% and a withdrawal speed of 33 µm s$^{-1}$. Faster speeds lead to lower coverage, thus less conductivity, e.g. at 66 µm s$^{-1}$ the resistivity increases to about 200 Ω. Speeds lower than 33 µm s$^{-1}$ do not increase the conductivity of the sample any further, indicating some sort of saturation in AgNW coverage. It was shown before on wood pulp fiber networks (paper) that the AgNW coating is proportional to the AgNW concentration in the suspension and the dip-coating parameters (Kopeinik 2015). An approximately linear relationship between withdrawal speed and conductivity was observed. The minimum coverage threshold value for noticeable conductivity of the viscose fiber is given by the fact that the AgNW need to form an interconnected network along the fiber. The actual amount of AgNW at this threshold value was not determined in this study. The electrical properties of the viscose/AgNW electrode fibers show perfectly ohmic behaviour, as demonstrated in an exemplary current-voltage diagram of a measured fiber in Figure 4. In total, 11 fibers were compared regarding their conductivity and an average resistivity of 45 ± 21 Ω was found.

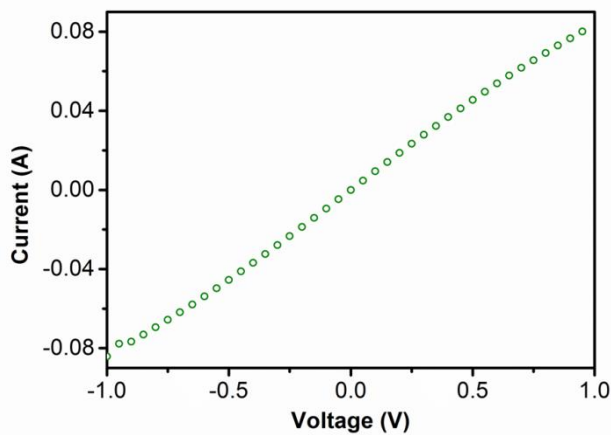

Figure 4: Current voltage behavior of a cellulose fiber dip-coated with silver nanowires.

The good adherence of silver nanowires to cellulosic surfaces has already been reported in (Kopeinik et al. 2015) on AgNW-decorated wood-pulp fibers. There, based on evidence from infrared spectroscopy it was suggested that the cellulose surface strongly interacts attractively with the silver nanowire surface. In addition, according electron microscopy images also showed a close alignment of the silver nanowires along the wood pulp fiber surface (Kopeinik et al. 2015) similar to the effect seen for viscose fibers in the SEM image in Figure 3. Therefore, we assume a similarly strong interaction for the silver nanowire adsorption to viscose fibers in the present case.

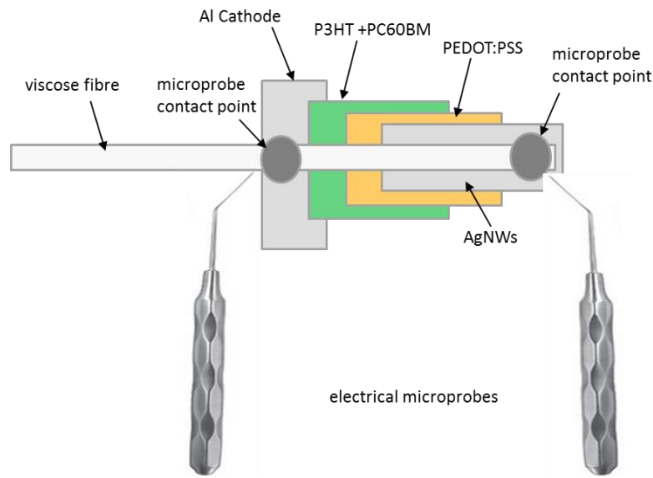

Figure 5: Sketch of the lineup of the different layers and contact points for the microprobes for electrical measurements.

Due to practical reasons for electrical measurement (avoid short-circuiting between the core and outer electrode during contacting) the precise design of the present viscose-based photovoltaic fibers was engineered in an overlapping structure, as shown in a schematic picture in Figure 5 and described as follows: The viscose fiber, there indicated by the white region, is not covered with AgNWs on the entire length, but only to one third (in the laboratory test device about 1 cm), represented by the light grey area surrounding the right-hand part of the fiber in the sketch. The PEDOT:PSS hole-conductor coating, indicated in yellow, leaves some part (2 mm) of the AgNW area free (on the right of the sketch) but covers a (2 mm) wider part of the viscose fiber, beyond the extend of the AgNW-coated area on the other end (on the left of the sketch). The subsequently applied organic semiconductor coating, indicated by the green area, is again shifted (1 mm) to the left, starting behind the PEDOT:PSS layer on one end, but extending beyond it on the other. Finally, the outer aluminum electrode is again deposited shifted (1mm) to the left, indicated in light grey on the left side of the sketch.

With this design, the aluminum coating only partially overlaps with the AgNW region underneath the semi-/hole-conductor layer of the photovoltaic fiber. Safe electrical contact to the nanowire network thus can be established on the bare AgNW end of the viscose fiber (indicated by the dark grey spot and the microprobe needle on the right hand side of the sketch). Safe contact to the outer aluminum electrode is made on that part of the fiber which is not overlapping with the AgNW region (indicated by the dark grey spot and the microprobe needle on the left hand side of the sketch). This leads to a total device length of about 15 mm. The overlap has to be chosen wisely, because if the distance for the charges generated in the semiconductor to the respective electrodes is too far, this is disadvantageous for the performance of the photovoltaic cell, due to the higher recombination probability of electron-hole pairs leading to low efficiencies.

The thickness of the organic hole-/semi-conductor coating on the photovoltaic fiber device is with about 1µm relatively large, compared to planar devices with usually 100-200 nm, but was required here to effectively cover the AgNW network. The transmission electron microscopy images in Figure 6 show exemplarily cross-sections of a sliced fiber with the coating configuration viscose/AgNW/P3HT:PCBM/gold. PEDOT:PSS and Al have been disregarded for facilitated TEM preparation.

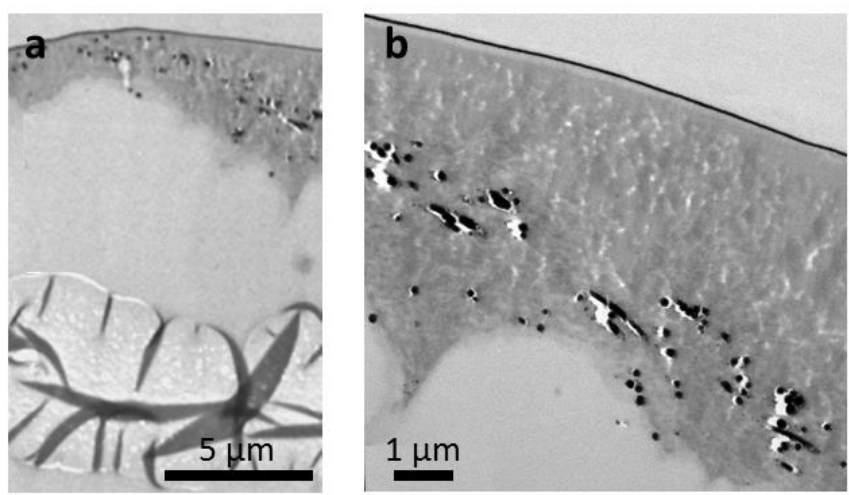

Figure 6: TEM images of a cross section of a photovoltaic fiber, recorded on a sliced fiber with the coating configuration viscose/AgNW/P3HT:PCBM/gold. (a) Collapsed viscose fiber (bottom of the image) and lifted-off AgNW/organic coating (top of the image) still reflecting the cellulose fiber's outer shape. (b) Magnification of the photovoltaic fiber's AgNW/organic/gold mantle. The cut AgNWs show as dark spots in the brighter organic matrix, the darker outer layer is the conductive gold coating.

In Figure 6a the collapsed viscose fiber is clearly visible in the lower half of the image. Above one can see the detached coating, which has been lifted off from the fiber by vacuum. The AgNWs are visible in the cross-section as dark spots within the brighter organic semiconductor matrix and on top a thin dark layer originating from the TEM gold coating. The detached coating reflects clearly the outer surface structure of the original viscose surface, like a finger-print. It is also visible from Figure 6a that the coating is not homogeneous, but varies in thickness between 0.7 and 5 µm due to the folds of the collapsed viscose fiber, leading to higher deposition in "valleys" than on higher regions. Such a thicker coating from a trench of a viscose fiber is seen as magnified cross-section in Figure 6b. Also here the surface topology of the viscose fiber is still preserved, visible in the former viscose-organic contact area at the bottom of the coating. Further, one can see that the distribution of silver nanowires in the coating, visible as dark spots, is restricted to a certain height in the mantle, indicating that they do not stick out or float when follow-on liquid materials are applied. This suggests that a homogeneous coating of a specific height around 1 µm should be sufficient to reliably cover the AgNWs and achieve reproducible fiber devices, even more on non-collapsing viscose fibers.

To determine the electrical device properties, current-voltage (IV) characteristics measurements of the photovoltaic fibers were recorded in the dark and under 0.54 mW/cm$^2$ LED white light illumination in inert atmosphere.

Exemplarily, the dark and photocurrent characteristics obtained for one photovoltaic fiber are shown in Figure 7 in one semi-logarithmic plot to enhance visibility. On first sight, both, dark current (represented by the black squares) and photocurrent (represented by the red circles) curves look quite symmetric, indicating ohmic behaviour caused by a considerable contribution of leakage currents in the order of $2 \cdot 10^{-5}$ A at zero bias. These are most likely caused by thickness inhomogeneities in the semiconductor coating on the collapsed viscose fiber surface, as mentioned above. However, a detailed analysis reveals that the devices still exhibit reasonable contributions of diode behavior, visible as a subtle difference between forward and reverse bias current, seen in the dark, but even more pronounced in the photo-IV plot. But even more important is the fact that the photovoltaic fiber shows indeed signs of light-induced current, visible as an open-circuit voltage ($V_{OC}$) different from zero, which would not be observed in a shortened cell, because of a total voltage drop at the shorts. Effectively, the presented exemplary photovoltaic fiber shows a $V_{OC}$ of 0.08 V and a short-circuit current ($J_{SC}$) of -0.5 mA/cm$^2$, leading to a value of 0.29 for the fill factor (FF). Taking the LEDs' light intensity into consideration, this results in a power conversion efficiency η of 0.023%. For calculation of current density, the photoactive device area was estimated with 0.025 cm$^2$, which is half of the viscose fiber's surface area (using the mean fiber diameter and the length of the device). Obtaining such solar cell key values for a not yet optimized photovoltaic viscose fiber device and at relatively low light intensity is quite satisfactory. For comparison, optimized planar devices of this configuration deposited on ITO glass usually deliver $V_{OC}$'s of max. 0.66 V, $J_{SC}$ between 8-12 mA/cm$^2$ and FF's between 0.50-0.65, under simulated solar conditions (AM1.5G and 100 mW/cm$^2$) (Dang, Hirsch, Wantz 2011), leading to power conversion efficiencies η of around 5%. We note that in the present case direct comparison of the dark and photocurrent of the fiber devices is difficult, due to instabilities of electrical contact when touching the quite delicate fiber with the comparatively macroscopic microprobes. In consequence there are inevitable contact resistance changes; accordingly the expected shift in reverse-bias current due to photogenerated charges is not apparent in the plot in Figure 7.

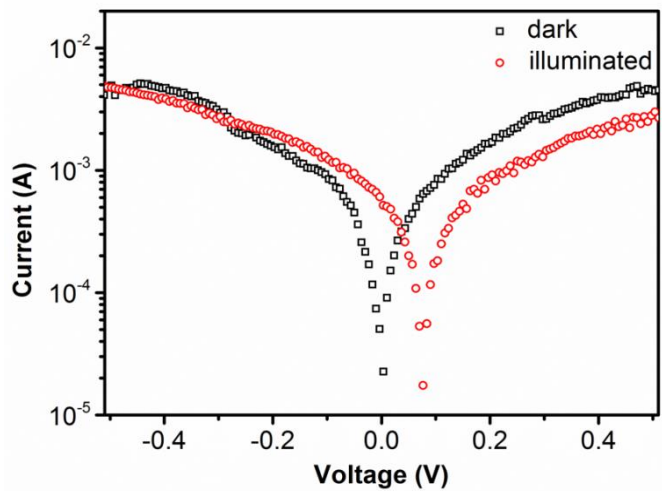

Figure 7: Current-voltage diagram of an illuminated (red circles) and dark (black squares) single cellulose fiber photovoltaic cell.

For a single cellulose fiber photovoltaic cell as a proof of concept, the results are remarkable. However, one clear culprit is the inhomogeneity of the semiconductor coating due to the collapse of the viscose fiber structure, leading to losses by not insignificant dark current contributions. A stable viscose core fiber should solve that problem and make way for advancement of the suggested concept for R2R production of photovoltaic fibers based on a common industrial cellulose fiber.

## Conclusions

In this paper, a proof-of-concept is given, showing that the fabrication of functional photovoltaic fibers based on common industrial cellulose fiber, is feasible. This has been demonstrated by preparation of short few centimeter long prototype photovoltaic fibers, built on commercially available hollow viscose fibers. Thereby the application of a silver nanowire surface network on the fiber forms the core electrode. This is followed by coatings of PEDOT:PSS for the hole-conductor and P3HT:PCBM for the photoactive semiconductor layer, all materials that are frequently used in common planar organic solar cells. The presented photovoltaic fibers show quite remarkable characteristics with fill factors and short-circuit current densities of comparable dimension to planar devices of this configuration, taking the relatively low light intensity during testing into account. Merely the open circuit voltage shows a considerable weakness, losses due to considerable amounts of leakage current as consequence of inhomogeneous semiconductor surface coverage, caused by the collapse of the chosen hollow viscose fiber. However, this obstacle can be solved by the choice of a viscose fiber with a flat, compact or reinforced structure. The stable dense structure of closely adhered silver nanowires on the viscose fiber surface enable reproducible core electrodes enable reproducible uniform follow-on coatings with a semiconductor mantle of limited thickness. The presented concept can be adapted easily to large scale role-to-role processing because all used materials are solution-processible and can be applied by dip-coating, spraying or printing. Fabrics made of such photovoltaic fibers could be integrated into smart clothing to generate the power for many wearable electronic devices.

Acknowledgements The Authors are indebted to Ingo Bernt and Walter Roggenstein from Kelheim Fibres GmbH for providing the fibers, the electron microscopy image of the fibers, and for discussions about the project.